\documentstyle[12pt,thmsa,sw20lart]{article}

\input{tcilatex}
\begin{document}

\title{Sufficient conditions of entanglement for tripartite and higher dimensional
multipartite qubit density matrixes}
\author{Zai-Zhe Zhong \\
Department of Physics, Liaoning Normal University, \\
Dalian 116029, Liaoning, China. E-mail: zhongzaizheh@hotmail.com}
\maketitle

\begin{abstract}
In this paper we give the new sufficient conditions of entanglement for
multipartite qubit density matrixes. We discuss in detail the case for
tripartite qubit density matrixes. As a criterion in concrete application,
its steps are quite simple and easy to operate. Some examples and
discussions are given.

PACC numbers: 03.67.Mn, 03.65.Ud, 03.67.Hk.
\end{abstract}

It is known that in modern quantum mechanics, especially in the quantum
information theory, to find the criteria of separability of density matrixes
is an important task. The first important result is the well-known positive
partial transposition(PPT, Peres-Horodecki) criteria[1,2$]$ for $2\times 2$
and $2\times 3$ systems. Some related works, see [3-7$].$ About the criteria
of separability for multipartite systems see [8-15$]$, especially a well
known result is the work of Rudolph[16]. As the simplest case of
multipartite systems, the criteria of separability of tripartite qubit
density matrixes are yet studied, e.g. see [$10,12$], and the classification
problem of tripartite qubit pure-states has been completed[17$]$. However,
for multipartite qubit systems, the problems are very complex, we always
hope to find some ways, which should be more simple and easy to operate, to
ascertain the existence of entanglement for multipartite qubit density
matrixes, even if they only are sufficient conditions. Since there always
are many known results about the separability for bipartite qubit density
matrixes, then we should consider such a problem: Can the problems of
multipartite qubit density matrixes be reduced, in some cases, to the
problems for several bipartite qubit density matrixes? If we can accomplish
this point, then we can ascertain the existence of entanglement for
multipartite qubit density matrixes with the aid of various known results
about bipartite qubit density matrixes. In this paper we give such a way,
though we only give some sufficient conditions, by which we can ascertain
the existence of entanglement in most cases.

The tripartite qubit states are most simple and important multipartite
states for application, hence in the first place, we study in detail the
case for tripartite qubit density matrixes. It is shown that the sufficient
condition of entanglement of a tripartite qubit density matrix is that any
one of several special reduced bipartite qubit density matrixes is
entangled. In concrete application, its steps are quite simple and easy to
operate. Some examples, discussions about conditions in theorem and
corollaries, and about the problem of generalization to more high
dimensional multipartite qubit density matrixes are given.

In this paper, we shall repeatedly use the definition of the separability of
a (bipartite or multipartite) state $\rho ,$ i.e. $\rho $ is separable if
and only if $\rho $ can be expresses by a convex sum that $\rho
=\sum\limits_\alpha p_\alpha \rho _\alpha ,0\leq p_\alpha \leq
1,\sum\limits_\alpha p_\alpha =1,$ where every $\rho _\alpha $ is a product
of the states of subsystems for any $\alpha $.

Let $\mid i_A>,\;\mid j_B$ \TEXTsymbol{>} and $\mid k_C>(i,j,k=0,1)$ be the
standard natural bases, they span the Hilbert spaces $H_A,\;H_B$ and $H_C$,
respectively. A tripartite qubit density matrix $\rho $ as an operator
acting upon $H_A\otimes H_B\otimes H_C$ can be written as 
\begin{equation}
\rho =\sum_{i,j,k,r,s,t=0}^1\left[ \rho \right] _{ijk,rst}\mid i_A>\mid
j_B>\mid k_C><r_A\mid <s_B\mid <t_C\mid
\end{equation}
where $\left[ \rho \right] _{ijk,rst}$ are the entries of matrix $\rho .$
Now we define six 4$\times 4$ matrixes $\rho _{\left( A,B\right) },\rho
_{\left( A,C\right) },\rho _{\left( B,C\right) },\rho _{\left( A,BC\right)
},\rho _{\left( B,CA\right) }$ and $\rho _{\left( C,AB\right) }$ as follows.
First, 
\begin{equation}
\rho _{\left( A,B\right) }=tr_C\left( \rho \right) ,\;\rho _{\left(
A,C\right) }=tr_B\left( \rho \right) ,\;\rho _{\left( B,C\right)
}=tr_A\left( \rho \right)
\end{equation}
Next, $\rho _{\left( A,BC\right) },\rho _{\left( B,CA\right) }$ and $\rho
_{\left( C,AB\right) }$, respectively, are defined by 
\begin{eqnarray}
\left[ \rho _{\left( A,BC\right) }\right] _{ij,rs} &=&\left[ \rho \right]
_{ijj,rss}+\left[ \rho \right] _{ij\left( 1-j\right) ,rs\left( 1-s\right) } 
\nonumber \\
\;\left[ \rho _{\left( B,CA\right) }\right] _{ij,rs} &=&\left[ \rho \right]
_{jij,srs}+\left[ \rho \right] _{\left( 1-j\right) ij,\left( 1-s\right) rs}
\\
\left[ \rho _{\left( C,AB\right) }\right] _{ij,rs} &=&\left[ \rho \right]
_{jji,ssr}+\left[ \rho \right] _{j\left( 1-j\right) i,s\left( 1-s\right) r} 
\nonumber
\end{eqnarray}

{\bf Lemma. }$\rho _{\left( A,B\right) },\rho _{\left( A,C\right) },\rho
_{\left( B,C\right) },\rho _{\left( A,BC\right) },\rho _{\left( B,CA\right)
} $ and $\rho _{\left( C,AB\right) }$ all are bipartite density matrixes.

{\bf Proof. }Since{\bf \ }$\rho _{\left( A,B\right) },\rho _{\left(
A,C\right) },\rho _{\left( B,C\right) }$ all are ordinary reductions of $%
\rho ,$ then lemma holds, we only need to make the proofs for $\rho _{\left(
A,BC\right) },\rho _{\left( B,CA\right) }$ and $\rho _{\left( C,AB\right) }$.

In the first place, we assume that $\rho $ is a pure-state, $\rho =\mid \Psi
><\Psi \mid ,$ where $\Psi =\sum\limits_{i,j,k=0}^1c_{ijk}\mid i_A>\mid
j_B>\mid k_C>$ $\in H_A\otimes H_B\otimes H_C$ is normalized, i.e. 
\begin{equation}
\sum\limits_{i,j,k=0,1}\left| c_{ijk}\right| ^2=1
\end{equation}
We take two form bases $\mid m_X>$ and $\mid n_Y>(m,n=0,1)$ and define 
\begin{eqnarray}
\Phi _{(A/BC)} &=&\left( \eta _{\left( A/BC\right) }\right)
^{-1}\sum_{m,n=0,1}c_{mnn}\mid m_X>\mid n_Y>  \nonumber \\
\;\Phi _{(A/B\stackrel{\vee }{C})} &=&\left( \eta _{(A/B\stackrel{\vee }{C}%
)}\right) ^{-1}\sum_{m,n=0,1}c_{mn\left( 1-n\right) }\mid m_X>\mid n_Y>
\end{eqnarray}
where $\eta _{\left( A/BC\right) }=\sqrt{\sum\limits_{m,n=0,1}\left|
c_{mnn}\right| ^2},\;\eta _{(A/B\stackrel{\vee }{C})}=\sqrt{%
\sum\limits_{m,n=0,1}\left| c_{mn\left( 1-n\right) }\right| ^2}$, then $\Phi
_{(A/BC)}$ and $\Phi _{(A/B\stackrel{\vee }{C})}$ both are normalized
bipartite qubit pure states. From Eqs.(4) and (5) we have 
\begin{equation}
\rho _{\left( A,BC\right) }=\eta _{\left( A/BC\right) }^2\rho
_{(A/BC)}+\;\eta _{(A/B\stackrel{\vee }{C})}^2\rho _{(A/B\stackrel{\vee }{C}%
)},\;\eta _{\left( A/BC\right) }^2+\eta _{(A/B\stackrel{\vee }{C})}^2=1
\end{equation}
where $\rho _{(A/BC)}=\mid \Phi _{(A/BC)}><\Phi _{(A/BC)},\rho _{(A/B%
\stackrel{\vee }{C})}=\mid \Phi _{(A/B\stackrel{\vee }{C})}><\Phi _{(A/B%
\stackrel{\vee }{C})}\mid .$ Since $\rho _{(A/BC)}$ and $\rho _{(A/B%
\stackrel{\vee }{C})}$ both are bipartite pure-states, $\rho _{\left(
A,BC\right) }$ is a bipartite density matrix (a mixed-state).

Secondly, if $\rho $ is a mixed-state and the expression $\rho
=\sum\limits_\alpha p_\alpha \rho _\alpha $ has been given, where every $%
\rho _\alpha $ is a pure-state with probability $p_\alpha $, then we can
obtain $\eta _{\alpha \left( A,BC\right) },\rho _{\alpha (A/BC)},\eta
_{\alpha (A,B\stackrel{\vee }{C})}$ and $\rho _{\alpha (A/B\stackrel{\vee }{C%
})}$ for every $\rho _\alpha $ . From Eq.(5), we have 
\begin{equation}
\rho _{\left( A,BC\right) }=\sum_\alpha \left( \zeta _{\alpha (A/BC)}\rho
_{\alpha (A/BC)}+\zeta _{\alpha (A/B\stackrel{\vee }{C})}\rho _{\alpha (A/B%
\stackrel{\vee }{C})}\right)
\end{equation}
where $\zeta _{\alpha (A/BC)}=p_\alpha \eta _{\alpha \left( A/BC\right)
}^2,\zeta _{\alpha (A/B\stackrel{\vee }{C})}=p_\alpha \eta _{\alpha (A/B%
\stackrel{\vee }{C})}^2,\sum_\alpha \left( \zeta _{\alpha (A/BC)}+\zeta
_{\alpha (A/B\stackrel{\vee }{C})}\right) =1.$ Since all $\rho _{\alpha
(A/BC)},\rho _{\alpha (A/B\stackrel{\vee }{C})}$ are bipartite qubit density
matrixes, $\rho _{\left( A,BC\right) }$ is a bipartite qubit density matrix
(a mixed state). Similarly, we can prove the cases for $\rho _{\left(
B,CA\right) }$ and $\rho _{\left( C,AB\right) }.$ {\bf QED}

From the above lemma we know that $\rho _{\left( A,BC\right) },\rho _{\left(
B,CA\right) }$ and $\rho _{\left( C,AB\right) }$, in fact, are yet some
special reductions of $\rho .$

{\bf Theorem (criterion of entanglement).} If any one of six bipartite qubit
density matrixes $\rho _{\left( A,B\right) },\rho _{\left( A,C\right) },\rho
_{\left( B,C\right) },\rho _{\left( A,BC\right) },$ $\rho _{\left(
B,CA\right) }$ and $\rho _{\left( C,AB\right) }$ is entangled, then the
tripartite qubit density matrix $\rho $ is entangled.

{\bf Proof. }In order to prove this theorem,{\bf \ }we only need to prove
that if $\rho $ is separable , then six bipartite qubit density matrixes $%
\rho _{\left( A,B\right) },\rho _{\left( A,C\right) },\rho _{\left(
B,C\right) },\rho _{\left( A,BC\right) },$ $\rho _{\left( B,CA\right) }$ and 
$\rho _{\left( C,AB\right) }$ all must be separable.

{\bf \ }In the first place, we prove that it is true when $\rho $ is a
pure-state. Suppose that $\rho =\mid \Psi _A><\Psi _A\mid \otimes \mid \Psi
_B><\Psi _B\mid \otimes \mid \Psi _C><\Psi _C\mid ,$ where $\Psi _A\equiv
a_0\mid 0_A>+a_1\mid 1_A>,$ $\Psi _B\equiv b_0\mid 0_B>+b_1\mid 1_B>,$ $\Psi
_C\equiv c_0\mid 0_C>+c_1\mid 1_C>,$ and $\left| a_0\right| ^2+\left|
a_1\right| ^2=\left| b_0\right| ^2+\left| b_1\right| ^2=\left| c_0\right|
^2+\left| c_1\right| ^2=1,$ then $\rho _{\left( A,B\right) }=\rho _A\otimes
\rho _B,\rho _{\left( B,C\right) }=\rho _B\otimes \rho _C,\rho _{\left(
A,C\right) }=\rho _A\otimes \rho _C,$ i.e. $\rho _{\left( A,B\right) },\rho
_{\left( B,C\right) }$ and $\rho _{\left( A,C\right) }$ are separable. In
addition, by a directly calculation we find the following equations:

\begin{eqnarray}
\rho _{\left( A,BC\right) } &=&\rho _A\otimes \omega _{BC},\;\omega
_{BC}=\left[ 
\begin{array}{cc}
\left| b_0\right| ^2 & \gamma _Cb_0b_1^{*} \\ 
\gamma _Cb_0^{*}b_1 & \left| b_1\right| ^2
\end{array}
\right] ,\;\gamma _C=2\func{Re} \left( c_0c_1^{*}\right)  \nonumber \\
\rho _{\left( B,CA\right) } &=&\rho _B\otimes \omega _{CA},\;\omega
_{CA}=\left[ 
\begin{array}{cc}
\left| c_0\right| ^2 & \gamma _Ac_0c_1^{*} \\ 
\gamma _Ac_0^{*}c_1 & \left| c_1\right| ^2
\end{array}
\right] ,\;\gamma _A=2\func{Re} \left( a_0a_1^{*}\right) \\
\;\rho _{\left( C,AB\right) } &=&\rho _C\otimes \omega _{AB},\;\omega
_{AB}=\left[ 
\begin{array}{cc}
\left| a_0\right| ^2 & \gamma _Ba_0a_1^{*} \\ 
\gamma _Ba_0^{*}a_1 & \left| a_1\right| ^2
\end{array}
\right] ,\;\gamma _B=2\func{Re} \left( b_0b_1^{*}\right)  \nonumber
\end{eqnarray}
where $\func{Re}\left( z\right) $ is the real part of a complex number $z,$ $%
\func{Re}\left( z\right) =\frac 12\left( z+z^{*}\right) $. Obviously, $%
\omega _{BC},\omega _{CA}$ and $\omega _{AB}$ all are bipartite qubit
density matrixes. Therefore $\rho _{\left( A,BC\right) },$ $\rho _{\left(
B,CA\right) }$ and $\rho _{\left( C,AB\right) }$ all must be separable.

Next, we discuss the case of mixed states. Suppose that $\rho $ is separable
mixed-state, i.e. there is a decomposition of $\rho $ as $\rho
=\sum\limits_\alpha p_\alpha \rho _\alpha ,0\leq p_\alpha \leq
1,\sum\limits_\alpha p_\alpha =1,$ and every $\rho _\alpha ${\bf \ }is a
separable pure-state. Then from the above discussions, all $\left( \rho
_\alpha \right) _{\left( \bullet ,\bullet \right) },\left( \rho _\alpha
\right) _{\left( \bullet ,\bullet \bullet \right) }($ $\left( \bullet
,\bullet \bullet \right) =\left( A,BC\right) ,\left( B,CA\right) ,\left(
C,AB\right) $ and $\left( \bullet ,\bullet \right) =\left( A,B\right)
,\left( B,C\right) ,\left( C,A\right) )$ all are separable pure-states. From
the definitions of $\left( \rho _\alpha \right) _{\left( \bullet ,\bullet
\right) },\left( \rho _\alpha \right) _{\left( \bullet ,\bullet \bullet
\right) }$ as in Eqs.($2)$ and (3)$,$ obviously we can obtain the following
equations 
\begin{equation}
\rho _{\left( \bullet ,\bullet \bullet \right) }=\sum\limits_\alpha p_\alpha
\left( \rho _\alpha \right) _{\left( \bullet ,\bullet \bullet \right)
},\;\rho _{\left( \bullet ,\bullet \right) }=\sum\limits_\alpha p_\alpha
\left( \rho _\alpha \right) _{\left( \bullet ,\bullet \right) }
\end{equation}
where $\left( \bullet ,\bullet \bullet \right) =\left( A,BC\right) ,\left(
B,CA\right) ,\left( C,AB\right) $ and $\left( \bullet ,\bullet \right)
=\left( A,B\right) ,\left( B,C\right) ,\left( C,A\right) .$ This means that
the bipartite mixes-states $\rho _{\left( A,B\right) },\rho _{\left(
A,C\right) },\rho _{\left( B,C\right) },\rho _{\left( A,BC\right) },$ $\rho
_{\left( B,CA\right) }$ and $\rho _{\left( C,AB\right) }$ all must be
separable. Sum up, whether $\rho $ is pure or mixed, if $\rho $ is
separable, then bipartite $\rho _{\left( A,B\right) },\rho _{\left(
A,C\right) },\rho _{\left( B,C\right) },\rho _{\left( A,BC\right) },$ $\rho
_{\left( B,CA\right) }$ and $\rho _{\left( C,AB\right) }$ all must be
separable, in other words, if any one of $\rho _{\left( A,B\right) },\rho
_{\left( A,C\right) },\rho _{\left( B,C\right) },\rho _{\left( A,BC\right)
}, $ $\rho _{\left( B,CA\right) }$ and $\rho _{\left( C,AB\right) }$ is
entangled, then the tripartite qubit density matrix $\rho $ must be
entangled. {\bf QED}

This theorem gives us a practical criterion (however they only are
sufficient conditions, see below) for existence of entanglement of $\rho $,
its steps are quite simple and easy to operate. As for whether one of six
density matrix is or not entangled, which may be ascertained by any known
way, say, by the PPT(Peres-Horodecki) criteria[$1,2].$ Therefore, for
instance, if we find any one partial transposition of $\rho _{\left(
A,B\right) },\rho _{\left( A,C\right) },\rho _{\left( B,C\right) },\rho
_{\left( A,BC\right) },\rho _{\left( B,CA\right) }$ and $\rho _{\left(
C,AB\right) }$ has a negative eigenvalue, then $\rho $ must be entangled.

{\bf Example 1.} As a simple example of pure-state, we see the GHZ-state $%
\rho =\mid \phi ><\phi \mid ,\phi =\frac 1{\sqrt{2}}\left( \mid 0_A>\mid
0_B>\mid 0_C>+\mid 1_A>\mid 1_B>\mid 1_C>\right) .$ In this case $\rho
_{\left( A,B\right) }=\rho _{\left( A,C\right) }=\rho _{\left( B,C\right)
}=\frac 12\mid 0_X><0_X\mid \otimes \mid 0_Y><0_Y\mid +\frac 12\mid
1_X><1_X\mid \otimes \mid 1_Y><1_Y\mid ,$ they all are separable. However $%
\rho _{\left( A,BC\right) }=\rho _{\left( B,CA\right) }=\rho _{\left(
C,AB\right) }$ $=\frac 12\left( \mid 0_X>\mid 0_Y>+\mid 1_X>\mid 1_Y>\right)
\left( <0_X\mid <0_Y\mid +<1_X\mid <1_Y\mid \right) $, they all are
entangled Bell's state, this shows that the GHZ-state is entangled.

{\bf Example 2. }If $\rho $ is a $8\times 8$ matrix whose entries are
defined by 
\begin{equation}
\left[ \rho \right] _{ijk,rst}=xR_{ijk,rst}+\frac 18\left( 1-x\right) \delta
_{ir}\delta _{js}\delta _{kt}
\end{equation}
where the real variable 0$\leq x\leq 1,$ the nonvanishing entries of density
matrix $R$ are $\left[ R\right] _{010,010}=\left[ R\right] _{011,011}=\left[
R\right] _{100,100}=\left[ R\right] _{101,101}=\frac 14,\;\left[ R\right]
_{010,101}=\left[ R\right] _{011,100}=\left[ R\right] _{100,011}=\left[
R\right] _{100,101}=-\frac 14$ . It is easily verified that $\rho $ is a
tripartite qubit density matrix. Therefore 
\begin{equation}
\left[ \rho _{(A,BC)}\right] _{ij,rs}=xS_{ij,rs}+\frac 14\left( 1-x\right)
\delta _{ir}\delta _{js}
\end{equation}
where the nonvanishing entries of density matrix $S$ are $\left[ S\right]
_{01,01}=\left[ S\right] _{10,10}=-\left[ S\right] _{01,10}=-\left[ S\right]
_{10,01}=\frac 12,$ i.e. $\rho _{(A,BC)}$ just is the Werner state[$1,18].$
From [1$]$ we know that the partial transposition of $\rho _{(A,BC)}$ has
three equal eigenvalues $\frac 14\left( 1+x\right) ,$ the fourth eigenvalue
is $\frac 14\left( 1-3x\right) .$ Therefore when $x>\frac 13,$ $\rho
_{(A,BC)}$ must be entangled, this leads that $\rho $ must be entangled. If
we don't use the above way, this is not easily seen.

Generally, if a bipartite qubit entangled density matrix $R$ is given, we
can define the nonvanishing entries of $\rho $ by any one of the following
six ways: 
\begin{eqnarray}
(1)\;\left[ \rho \right] _{ijj,rss} &=&\left[ \rho \right] _{ij\left(
1-j\right) ,rs\left( 1-s\right) }=\frac 12\left[ R\right] _{ij,rs}  \nonumber
\\
\left( 2\right) \;\left[ \rho \right] _{jij,srs} &=&\left[ \rho \right]
_{\left( 1-j\right) ij,\left( 1-s\right) rs}=\frac 12\left[ R\right] _{ij,rs}
\nonumber \\
\left( \text{3}\right) \;\left[ \rho \right] _{jji,ssr} &=&\left[ \rho
\right] _{j\left( 1-j\right) i,s\left( 1-s\right) r}=\frac 12\left[ R\right]
_{ij,rs}\;  \nonumber \\
\left( \text{4}\right) \;\left[ \rho \right] _{ij0,rs0} &=&\left[ \rho
\right] _{ij1,rs1}=\frac 12\left[ R\right] _{ij,rs} \\
\left( \text{5}\right) \;\left[ \rho \right] _{i0j,r0s} &=&\left[ \rho
\right] _{i1j,r1s}=\frac 12\left[ R\right] _{ij,rs}\;  \nonumber \\
\left( \text{6}\right) \;\left[ \rho \right] _{0ij,0rs} &=&\left[ \rho
\right] _{1ij,1rs}=\frac 12\left[ R\right] _{ij,rs}  \nonumber
\end{eqnarray}
then the tripartite qubit density matrix $\rho $ must be entangled$.$

{\bf Example 3. }In [$19]$, a state considered (we only discuss the case of
tripartite qubit state) is as 
\begin{eqnarray}
\rho &=&\sum_{rs=AB,BC,AC}p_{rs}\mid \Psi _{rs}><\Psi _{rs}\mid  \nonumber \\
\; &\mid &\Psi _{rs}>=\frac 1{\sqrt{2}}\left( \mid 0_r>\mid 1_s>+\mid
1_r>\mid 0_s>\right) \otimes \mid 0_{rest}>
\end{eqnarray}
where $0\leq p_{rs}\leq 1,\;\sum\limits_{rs=AB,AC,BC}p_{rs}=1.$ In order to
clarify that whether $\rho $ is or not entangled, we consider $\rho _{\left(
r,s\right) }(\left( r,s\right) =\left( A,B\right) ,\left( A,C\right) ,\left(
B,C\right) ).$ It is proved[18] that the form of $\rho _{\left( r,s\right) }$
is as 
\begin{equation}
\rho _{\left( r,s\right) }=\left[ 
\begin{array}{cccc}
\alpha _{rs} &  &  &  \\ 
& \beta _{rs} & \frac 12p_{rs} &  \\ 
& \frac 12p_{rs} & \gamma _{rs} &  \\ 
&  &  & 0
\end{array}
\right]
\end{equation}
where $\alpha _{rs},\beta _{rs},\gamma _{rs}$ are some real numbers and $%
\alpha _{rs}+\beta _{rs}+\gamma _{rs}=1$. Here, there is no need to
calculate the so-called `concurrences'[20] of $\rho _{\left( r,s\right) }$
as did as in [19$],$ but we can use the above criteria. In fact, the partial
transposition of matrix $\rho _{\left( r,s\right) }$ has four eigenvalues $%
\lambda _{\left( 1\right) rs}=\beta _{rs},\lambda _{\left( 2\right)
rs}=\gamma _{rs},\lambda _{\left( \pm \right) rs}=\frac 12\left( -\alpha
_{rs}\pm \sqrt{\alpha _{rs}^2+p_{rs}^2}\right) .$ It is impossible that $%
p_{AB},p_{AC},p_{BC}$ all vanish simultaneously, this means that there must
be at least one of three $\lambda _{\left( -\right) rs}(rs=AB,AC,BC)$ which
is negative, therefore $\rho $ must be entangled. In view of physical point,
it is true of course, since the state has contained some `entanglement
molecules'[19$].$

{\bf Discussion.} (1) It is a pity that the conditions in theorem only are
sufficient, but not necessary condition for entanglement of $\rho .$ In
fact, for some $\rho $ the six reduced matrixes all are separable, however $%
\rho $ still are entangled. For instance, a state given in [$8$] is as 
\begin{equation}
\rho =\frac 14\left( I-\sum_{i=1}^4\mid \psi _i><\psi _i\mid \right)
\end{equation}
where $\mid \psi _1>=\frac 1{\sqrt{2}}\mid 0_A>\mid 1_B>\mid \left( \mid
0_C>+\mid 1_C>\right) ,\;\mid \psi _2>=\frac 1{\sqrt{2}}\mid 1_A>\left( \mid
0_B>+\mid 1_B>\right) \mid 0_C>,\mid \psi _3>=\frac 1{\sqrt{2}}\left( \mid
0_A>+\mid 1_A>\right) \mid 1_B>\mid 0_C>,$

$\mid \psi _4>=\frac 1{2\sqrt{2}}\left( \mid 0_A>-\mid 1_A>\right) \left(
\mid 0_B>-\mid 1_B>\right) \left( \mid 0_C>-\mid 1_C>\right) ,$ in this case 
$\rho _{\left( A,B\right) },\rho _{\left( A,C\right) },\rho _{\left(
B,C\right) },\rho _{\left( A,BC\right) },\rho _{\left( B,CA\right) }$ and $%
\rho _{\left( C,AB\right) }$ all are separable (it can be verified that they
all satisfy the PPT conditions), however $\rho $ is entangled[8,12].

Similarly, $\rho _{\left( AC,BD\right) },\rho _{\left( AD,BC\right) }$
(notice that there are several repeated, e.g. $\rho _{\left( BD,AC\right)
}=\rho _{\left( AC,BD\right) },$ etc., hence we delete them). By a similar
way we can define $\rho _{\left( A,BCD\right) }.\rho _{\left( B,CDA\right)
},\rho _{\left( C,DAB\right) },\rho _{\left( D,ABC\right) },$ etc.. We can
similarly prove that the above bipartite matrixes all are density matrixes,
and we have completely similar theorem, etc.. By the similar way, in fact,
we can obtain the more generalized results of separability and partial
separability (hence the entanglement and the partial entanglement) of
multipartite qubit density matrixes, they are discussed by us elsewhere[21].

\end{document}